\documentclass{ws-ijmpd}

\usepackage{graphicx}  
\usepackage{latexsym}
\usepackage{overcite}  

\def\beq{\begin{equation}}
\def\eeq{\end{equation}}
\def\rmd{{\rm d}}
\def\rmD{{\rm D}}
\def\version{\today}

\def\fl{}

\begin{document}

%
\def\nocropmarks{\vskip5pt\phantom{cropmarks}}
\let\trimmarks\nocropmarks      
%

\markboth{Bini D., Cherubini C., Geralico A. and Jantzen R.}
{Massless spinning test particles}

%
\catchline{}{}{}
%

\title{
MASSLESS SPINNING TEST PARTICLES IN ALGEBRAICALLY SPECIAL VACUUM SPACETIMES
}

\author{\footnotesize DONATO BINI}

\address{
Istituto per le Applicazioni del Calcolo ``M. Picone'', CNR I--00161 Rome, Italy,  \\
ICRA,
University of Rome ``La Sapienza'', I--00185 Rome, Italy,\\
INFN, Sezione di Firenze, I-50019 Sesto Fiorentino (FI), Italy
\footnote{binid@icra.it}
}

\author{CHRISTIAN CHERUBINI}

\address{
Faculty of Engineering, University Campus Bio-Medico,
I--00155 Rome, Italy, \\
ICRA,
University of Rome ``La Sapienza'', I--00185 Rome, Italy
\footnote{cherubini@icra.it}
}

\author{ANDREA GERALICO}

\address{
ICRA,
University of Rome ``La Sapienza'', I-00185 Rome, Italy
\footnote{geralico@icra.it}
}

\author{ROBERT T JANTZEN}
\address{
Department of Mathematical Sciences, Villanova University, Villanova, PA, 19085, USA,\\
ICRA,
University of Rome ``La Sapienza'', I-00185 Rome, Italy
\footnote{robert.jantzen@villanova.edu}
}

\maketitle

\begin{history}
\received{\version}
\revised{}
\end{history}

\begin{abstract}
The motion of massless spinning test particles is investigated using the Newman-Penrose formalism within the Mathisson-Papapetrou model extended to massless particles by Mashhoon and supplemented by the Pirani condition. When the \lq\lq multipole reduction world line" lies along a principal null direction of an algebraically special vacuum spacetime, the equations of motion can be explicitly integrated. Examples are given for some familiar spacetimes of this type in the interest of shedding some light on the consequences of this model.
\end{abstract}


\section{Introduction}

The motion of a  massive spinning test particle in general relativity is described by the Mathisson-Papapetrou model  \cite{math37,papa51} consisting of 10 equations
\begin{eqnarray}
\label{papcoreqs1}
\frac{DP^\alpha}{\rmd\tau_U}&=-\frac12 R^\alpha{}_{\beta\rho\sigma}U^\beta S^{\rho\sigma}, \\
\label{papcoreqs2}
\frac{DS^{\alpha\beta}}{\rmd\tau_U}&=[P\wedge U]^{\alpha\beta},
\end{eqnarray}
for 13 unknowns: the generalized momentum of the particle $P^\alpha $ (4), the antisymmetric spin tensor $S^{\alpha\beta}$ (6) and the unit tangent vector $U^\alpha=\rmd x^\alpha/\rmd\tau_U$ to the world line used for the multipole reduction (3).

For the model to be complete, it must supplemented by three additional equations which in the literature are usually called \lq\lq supplementary conditions\rq\rq (SCs). There are three standard choices for the supplementary conditions:
\begin{enumerate}
\item[1.] Papapetrou-Corinaldesi \cite{cori51}: $S^{\alpha \beta}u_\beta=0$,
where $u$ is some arbitrary timelike vector chosen for convenience of calculation;
\item[2.] Pirani \cite{pir56}: $S^{\alpha\beta}U_\beta=0$;
\item[3.] Tulczyjew \cite{tulc59}: $S^{\alpha\beta}P_\beta=0$.
\end{enumerate}
Each of these is an algebraic condition restricting the spin tensor to have 3 fewer independent components. If the vector in terms of which the supplementary condition is expressed is timelike, this just means that the spin tensor is equivalent to a spin vector lying in the local rest space associated with this timelike direction. 
In a sense, only the Pirani and Tulczyjew SCs are \lq\lq intrinsic" relations between the various unknowns of the model and have more credibility as physical conditions rather than just a mathematical convenience which is  \lq\lq coordinate dependent", as is the Papapetrou-Corinaldes SC.

The extension of the Mathisson-Papapetrou model to the case of a null multipole reduction world line with tangent $l$ has been considered by Mashhoon \cite{mash75}.
The equations of the extended model have exactly the same form as (\ref{papcoreqs1}) and (\ref{papcoreqs2}) but with $U$ replaced by $l$ and $\tau_U$ (the proper time parameter along the timelike world line) replaced by $\lambda$ (a parameter along the null world line such that $d x^\alpha /d\lambda =l^\alpha$)
\begin{eqnarray}
\label{papcoreqs1n}
\frac{\rmD P^\alpha}{\rmd\lambda }&=-\frac12 R^\alpha{}_{\beta\rho\sigma}l^\beta S^{\rho\sigma}\ , \\
\label{papcoreqs2n}
\frac{\rmD S^{\alpha\beta}}{\rmd\lambda}&=[P\wedge l]^{\alpha\beta}\ .
\end{eqnarray}
Eqs.~(\ref{papcoreqs1n})  and (\ref{papcoreqs2n}) must then be solved assuming some SC.
Consider only the more physically motivated SCs of Pirani and Tulczyjew, with Pirani's conditions now naturally generalized to $S^{\alpha\beta}l_\beta=0$.
Furthermore, since  we are interested in the massless limit of the Mathisson-Papapetrou equations, and since the mass of the particle is defined by
$m=-P\cdot U$,
the massless limit would imply $P\cdot l =0$, so we assume this to be true here.

Mashhoon \cite{mash75} found Tulczyjew's condition
$S^{\alpha\beta}P_\beta=0$ to be inconsistent if $P$ is also
null: $P\cdot P=0$. 
While we could examine this condition for the nonnull case, 
it seems most reasonable to consider Pirani's SC as the only physically
meaningful supplementary condition. Under these conditions
Mashhoon \cite{mash75} has shown that the particle follows a null
geodesic (so that one can assume $\lambda$ to be an affine parameter)
with its  spin vector (denoted by ${\mathcal S}$)
 \beq 
 \label{spinvector} 
 {\mathcal S}{}^\mu=\frac12
\eta^{\mu\nu\alpha\beta}l_\nu S_{\alpha\beta} 
\eeq 
parallelly transported along the null world line with tangent
$l$ and either parallel or antiparallel to $l$. Moreover, again adopting 
Mashhoon's point of view, the generalized momentum of
the particle should be spacelike or even null and is
determined by solving equations (\ref{papcoreqs1n}) and
(\ref{papcoreqs2n}) supplemented by $S^{\alpha\beta}l_\beta=0$. The
other components of the spin tensor not equivalent to the spin
vector that one can introduce must also be 
determined.\footnote{It is worth noting that
the Mathisson-Papapetrou model does not place {\it a priori}
restrictions on the causal character of $U$ and $P$ and there is no
agreement in the literature on this freedom.
For example, Tod,  de Felice and Calvani \cite{fdf} consider $P$
to be timelike assuming that it  represents the total energy-momentum
of the particle while they do not impose any causality
condition on the world line $U$, which plays the role of a mere
mathematical \lq\lq tool" to perform the multipole reduction;
on the other hand according to Mashhoon \cite{mash75}, $P$ can be
considered to be analogous to a canonical momentum for the particle,
in which case there is no reason to prefer a particular causality character for it,
while the world line $U$ must be timelike in the massive case or null in the massless case
because it is a world line intimately associated with the test particle location. This
uncertainty in the model itself, i.e., the uncertainty in the SC, has
to be considered as a limitation of our description of spinning test
particles in general relativity. We are aware of this problem
and only consider the Pirani SC here.}
Note that the spin vector depends on the scale of $l$ and hence on the parametrization chosen for the null world lines to which it is tangent.

In the present paper, using the Newman-Penrose (NP) formalism,
we derive the equations of  motion for a massless spinning test particle in any algebraically special vacuum spacetime for the special case in which the multipole reduction world line is aligned with a principal null direction of the spacetime. This leads to very simple equations which are straightforward to solve.
Explicit solutions corresponding to some familiar  Petrov type D and type N spacetimes (including the Schwarzschild, Kerr, C metric, Kasner, and single exact gravitational wave spacetimes) are derived and discussed. Finally, motion along null circular orbits is also studied and explicit solutions in black hole spacetimes are presented.

We adopt the standard NP notation of Chandrasekhar \cite{ChandraBook} which includes the metric signature $+$ $-$ $-$ $-$.

\section{Equations of motion in the Newman-Penrose formalism}

In this section we express the above equations for massless spinning test particles in the NP formalism.
Let  $e_1=l$, $e_2=n$, $e_3=m$, $e_4=\bar m$ be an  NP frame consisting of four null vector fields satisfying the relations
$l \cdot n=1$ and $m \cdot \bar m=-1$, so that
the frame components of the metric tensor are
\beq
(\eta_{ab}) = (\eta^{ab})
=\pmatrix{
0 & 1 & 0 & 0 \cr
1 & 0 & 0 & 0 \cr
0  & 0 & 0 & -1\cr
0  & 0 & -1& 0 \cr
}.
\eeq
Exactly like the NP representation of the electromagnetic Faraday 2-form,
the 6 real components of the spin tensor $S$ are represented in terms of 3 complex scalar quantities $S_0,S_1,S_2$
\beq
(S_{ab}) = 
\pmatrix{
0 & \bar S_1+S_1 & S_0 & \bar S_0 \cr
-\bar S_1-S_1 & 0 & -\bar S_2 &- S_2 \cr
-S_0 & \bar S_2 & 0 & \bar S_1-S_1\cr
-\bar S_0 & S_2 &  -\bar S_1+S_1 &0\cr
}\, ,
\eeq
which have the well-defined spin-weights \cite{ghp} $-1,0,1$ respectively.\footnote{An NP frame aligned with a given null direction $l$ is only fixed up to boosts and rotations which leave the directions of $l$ and $n$ unchanged, i.e., up to class III tetrad Lorentz transformations
$$
l^\mu \to A l^\mu, \quad n^\mu \to A^{-1}n^\mu, \quad m^\mu \to e^{i\theta}m^\mu\ .
$$
A quantity $\zeta$ which tranforms under this transformation as
$$
\zeta \to A^{(p+q)/2}e^{i(p-q)/2}\zeta\ ,
$$
is said to be a quantity of type $(p,q)$, with spin weight $s=(p-q)/2$ and boost weight $w=(p+q)/2$. Most of the NP quantities have a well-defined spin and boost weight. The exceptions are the spin coefficients $\alpha, \beta, \gamma, \epsilon$ and the directional derivative operators $D,\Delta, \delta, \bar \delta$  along the frame vectors.
}

We first adapt the NP frame to the null world line of the multipole reduction by choosing its tangent as the first frame vector: $l=e_1$. The zero mass condition $0=P\cdot l = P^2$ then forces the second component $P^2$ of $P$ to be zero  (and hence $P\cdot P=-2|P^3|^2$, so $P$ is spacelike).
Eqs.~(\ref{papcoreqs1n}) and (\ref{papcoreqs2n}) then take the following form in a vacuum spacetime where the Riemann curvature tensor reduces to the Weyl curvature tensor (NP scalars $\psi_0,\psi_1,\psi_2,\psi_3,\psi_4$)
\begin{eqnarray}
\fl
\frac{\rmd S_0}{\rmd \lambda} &=& -2 \kappa S_1+ 2\epsilon S_0\ ,\quad
\fl
\frac{\rmd S_1}{\rmd \lambda} = \pi S_0- \kappa S_2\ ,\quad
\fl
\frac{\rmd S_2}{\rmd \lambda} = 2\pi S_1-2 \epsilon S_2-P^3\ ,\nonumber \\
\fl
\frac{\rmd P^1}{\rmd \lambda} &=& -(\epsilon+\bar \epsilon )P^1 -\pi\bar P^3 - \bar \pi P^3 
- \psi_3 S_0 -\bar \psi_3 \bar S_0 
+ 2 \psi_2 S_1 + 2 \bar \psi_2 \bar S_1 
- \psi_1 S_2 - \bar \psi_1 \bar S_2\ ,\nonumber \\
\fl
\frac{\rmd P^3}{\rmd \lambda} &=&   \bar \kappa P^1 +(\bar \epsilon-\epsilon )P^3 
   -2 \bar \psi_1 \bar S_1 + \bar \psi_2 \bar S_0+ \bar \psi_0 \bar S_2\ ,\quad
0 =  \bar\kappa \bar P^3+\kappa P^3\ ,
\end{eqnarray}
where $\rmD/\rmd\lambda=l^{\alpha}\nabla_{\alpha}\equiv D$ is the directional derivative along $l$. As usual the equation for $P^4$ is just the complex conjugate of the one for $P^3$.

Pirani's supplementary conditions 
are simply:\footnote{Mashhoon's notation used in \cite{mash75} corresponds to $f=S_2$ and $g=s$ and his choice of frame transport law along $l$ is  equivalent to the conditions $\pi=0$ and $\epsilon + \bar \epsilon=0$ on the NP scalars which guarantee that  $l$ is the tangent to an affinely parametrized geodesic and that $n$ is parallel transported along $l$. This latter assumption is unnecessary here.}
\begin{eqnarray}
(S_1+\bar S_1 ) =0\ , \quad  S_0 =0\ ,
\end{eqnarray}
i.e., $S_1=is/2$ is purely imaginary with $s$ real and whose meaning is discussed below (don't confuse $s$ with the symbol for the spin-weight of tensorial quantities).
The only nonvanishing frame components of the spin tensor are then 
$S_{24}=-S_{2}$ and $S_{34}=-is$.
With these conditions the equations of motion then become
\begin{eqnarray}
\label{eqsspinP}
\fl 0&=& \kappa s\ ,\quad
\fl \frac{\rmd s}{\rmd \lambda} = i 2\kappa S_2\ ,\quad
\fl \frac{\rmd S_2}{\rmd \lambda}  = \pi is -2 \epsilon S_2-P^3\ ,\nonumber \\
\fl \frac{\rmd P^1}{\rmd \lambda} &=& -(\epsilon+\bar \epsilon )P^1 -\pi\bar P^3 - \bar \pi P^3  
   +i  s\, (\psi_2-\bar \psi_2)- \psi_1S_2- \bar \psi_1 \bar S_2\ ,\nonumber \\
\fl 0
 &=& \bar\kappa \bar P^3+\kappa P^3\ ,\quad
\fl \frac{\rmd P^3}{\rmd \lambda}  =   \bar \kappa P^1 +(\bar \epsilon-\epsilon )P^3
   +i \bar \psi_1 s+ \bar \psi_0 \bar S_2\ .
\end{eqnarray}
The solution $s=0$ of the first equation would lead to $S_2=0$ and hence vanishing spin tensor, a trivial case that will not be considered here. One must therefore have instead $\kappa=0$, i.e., $l$ is a null geodesic \cite{ChandraBook}. Hence, these equations further reduce to
\begin{eqnarray}
\label{eqsspinP_fin}
\fl \frac{\rmd s}{\rmd \lambda} &=&0\ ,\quad
\fl \frac{\rmd S_2}{\rmd \lambda}  = \pi is -2 \epsilon S_2-P^3\ ,\nonumber \\
\fl \frac{\rmd P^1}{\rmd \lambda} &=& -(\epsilon+\bar \epsilon )P^1 -\pi\bar P^3 - \bar \pi P^3  +i  s\, (\psi_2-\bar \psi_2)- \psi_1S_2- \bar \psi_1 \bar S_2\ ,\nonumber \\
\fl \frac{\rmd P^3}{\rmd \lambda}  &=& (\bar \epsilon-\epsilon )P^3+i \bar \psi_1 s+ \bar \psi_0 \bar S_2\ .
\end{eqnarray} 
A further simplification can be made by using a type III rotation of the 
NP frame 
to set the spin coefficient $\epsilon=0$. This corresponds to require 
that the 
congruence of $l$ is affinely parametrized. 
The final set of equation is the following: 
\begin{eqnarray} 
\label{eqsspinP_fin2} 
\frac{\rmd s}{\rmd \lambda} &=&0\ , \qquad 
\frac{\rmd S_2}{\rmd \lambda}  = \pi is -P^3\ ,\nonumber \\ 
\frac{\rmd P^1}{\rmd \lambda} &=& -\pi\bar P^3 - \bar \pi P^3  +i  s\, 
(\psi_2-\bar \psi_2)- \psi_1S_2- \bar \psi_1 \bar S_2\ ,\nonumber \\ 
\frac{\rmd P^3}{\rmd \lambda}  &=& i \bar \psi_1 s+ \bar \psi_0 \bar S_2\ . 
\end{eqnarray} 
These equations are still general for any vacuum spacetime. 
In the special case that the 4-momentum is a null vector, equivalent to 
$P^3=0$ as seen above, then $P=P^1 \, l$ and the previous set of 
equations specialize to 
\begin{eqnarray} 
\label{eqsspinPP3eq0} 
\frac{\rmd s}{\rmd \lambda} & =& 0\ , \qquad \frac{\rmd S_2}{\rmd \lambda} 
= \pi is\ ,\nonumber \\ 
\frac{\rmd P^1}{\rmd \lambda} & =& i  s\, (\psi_2-\bar \psi_2)- 
\psi_1S_2- \bar \psi_1 \bar S_2\ , 
\qquad 0= -i \psi_1 s + \psi_0  S_2\ . 
\end{eqnarray}

\section{Interpreting the spin components}

The spin vector defined in Eq.~(\ref{spinvector}) as the contracted dual of the spin tensor 
\beq
\mathcal{S}^\mu = {}^* S^{\mu\nu} l_\nu \quad
\rightarrow \quad \mathcal{S} = i(\bar S_1 - S_1) l + i S_0 m - i \bar S_0 \bar m\ ,
\eeq
when the Pirani supplementary conditions $S_1=is/2$ and $S_0=0$ are imposed,
reduces to the simple form
\beq \label{spinvector_new}
{\mathcal S}^\mu = s \, l^\mu\ ,
\eeq 
so that $s$  represents the helicity of the particle (see p.~80 of reference \citen{RP} for the relation between the helicity and spin-weight of a massless field). Since $s$ is constant along the world line, and $l$ is parallelly transported along itself, the spin vector is parallel transported along $l$, recovering Mashhoon's result.

However, the
spin tensor is not completely determined by the spin vector alone as in the timelike case but also requires the independent component $S_2$
\beq \label{SNP} 
S=l \wedge (S_2 m + \bar S_2 \bar m) + is\, m \wedge \bar m\ .
\eeq 
The real vector $(S_2 m + \bar S_2 \bar m)$ can be interpreted as a sort of \lq\lq orbital angular momentum" contribution to the total spin (see reference \citen{RP}, p.~68),
as measured with respect to the null world line of the multipole
reduction associated with $l$. In a sense, 
since $l$ is a null vector, the multipole moments taken with
respect to any spacetime point on the null world line are always those of a particle in motion with respect to any physical observer. 

Alternatively, there are no observers  such that  
the associated electric part of $S$ vanishes identically as it does in the timelike case. 
To see this let us consider the Lorentz frame naturally associated with the
NP frame, defined by
\begin{eqnarray}
\fl \quad 
l=\frac{1}{\sqrt{2}}(E_0+E_1)\ , \quad
n=\frac{1}{\sqrt{2}}(E_0-E_1)\ , \quad 
m=\frac{1}{\sqrt{2}}(E_2+iE_3)\ .
\end{eqnarray}
Using the notation $U=E_0$.
Eq.~(\ref{SNP}) then becomes
\beq
S = U\wedge L(U)+{}^{*(U)}S(U)\ ,
\eeq
where
\begin{eqnarray}
\label{elemag}
L(U)&=& {\rm Re}\, [S_2] E_2 -{\rm Im}\, [S_2] E_3\ , \nonumber \\
S(U)&=& {\rm Re}\, [S_2] E_3 +{\rm Im}\, [S_2] E_2 +s\, E_1\ ,
\end{eqnarray}
which are respectively the electric and magnetic parts of $S$ with respect to $U$.
The notation for the spatial dual of a spatial vector $X$ (spatial with respect to $U$, i.e., $X\cdot U=0$), 
\beq
[{}^{*(U)}X]^{\alpha\beta}=\frac12 \eta^{\mu \nu \alpha \beta}U_\mu X_\nu\ ,
\eeq
as well as for the exterior product of two spatial vectors $X$ and $Y$ (spatial with respect to $U$) 
\beq
[X\times_U Y]^\alpha = \eta^{\beta\alpha\mu\nu}U_\beta X_\mu Y_\nu\ ,
\eeq
where $\eta_{\alpha\beta\gamma\delta}$ is the unit volume four form,  will be used here and below \cite{mfg}.
Note that
\beq
L(U)+iS(U)=\sqrt{2}S_2 \, m +is E_1\ .
\eeq

The two algebraical invariants of $S$ can then be written as 
\beq 
\fl
I_1 = \frac12 S^{\alpha\beta}S_{\alpha\beta}
    = -L(U)^2+S(U)^2\equiv s^2, \quad 
I_2 = \frac12 S^{\alpha\beta}{}^*S_{\alpha\beta}
    =L (U) \cdot S(U)\equiv 0\ . 
\eeq 
The spin tensor can be classified in the same way as the electromagnetic 2-form. If the (complex) invariant
\beq
\fl\quad
I = \frac14 (S^{\alpha\beta}+i{}^*S^{\alpha\beta})(S_{\alpha\beta}+i{}^*S_{\alpha\beta})
  = I_1+iI_2 = 4(S_0 S_2 - S_1^2)
\eeq
is nonzero, the spin tensor will be said to be non-null or nonsingular, while if it is zero the spin tensor will be said to be  null or singular.
In the present case since $I_2=0$, it reduces to $I=s^2$ and the spin tensor is singular only when $s=0$. Note that the spinning test
particle model holds when a dimensionless quantity formed from $s$ is very small: otherwise modification of the background due to backreaction should be considered and the particle cannot be considered to be a test particle. 
Due to the choice of P supplementary conditions, the quantity $S_2$ does not contribute to the spin invariant; therefore, requiring the magnitude of $s$ to be small does not give any restriction on it. 

The 4-velocity $u$ of any physical observer can be expressed in terms of its relative velocity with respect to $U$ and vice versa \cite{mfg}
\beq
u=\gamma(U,u)[U+\nu(u,U)]\ ,\qquad 
U=\gamma (U,u)[u+\nu(U,u)]\ ,
\eeq
where
\beq
\fl\quad
||\nu(U,u)||^2 = ||\nu(u,U)||^2\ , \quad 
\gamma(U,u) = 1/\sqrt{1-||\nu(U,u)||^2} = \gamma(u,U)\ .
\eeq
Then 
\begin{eqnarray}
S&= u\wedge L(u)+{}^{*(u)}S(u)\ ,
\end{eqnarray}
with $L(u)$ and $S(u)$ related to $L(U)$ and $S(U)$ by a boost transformation exactly like the electric and magnetic fields \cite{mfg}
\begin{eqnarray}
\label{em_transf}
P(U)L(u)&= \gamma (U,u) [L(U)+\nu(u,U)\times_U S(U)]\ ,\nonumber \\
P(U)S(u)&= \gamma (U,u) [S(U)+\nu(u,U)\times_U L(U)]\ ,
\end{eqnarray}
where $P(U)=1+U\otimes U $ projects orthogonally to $U$.
Since $I_1=0$, it follows that $L(u)$ and $S(u)$
are orthogonal, i.e., any observer will see the electric and magnetic parts of the spin tensor to be orthogonal. Moreover, 
from (\ref{elemag}), it easy to see that 
\beq L(U)+E_1 \times S(U)=0\ ,
\eeq 
which compared with (\ref{em_transf}), gives $L(u)=0$ only for an observer moving with respect to $U$ with velocity 
$\nu(u,U)=E_1$, i.e., with unit speed, so that it does not correspond to any physical observer.

\section{Vacuum algebraically special spacetimes}

The Goldberg-Sachs theorem \cite{kraetal} states that a vacuum metric is algebraically special if and only if it contains a shearfree null geodesic congruence with tangent $l$ (taken as the first vector of an NP frame), i.e.,
\beq
\kappa=0=\sigma, \qquad \psi_0=0=\psi_1\ .
\eeq
Hence, assuming that the (single) line of multipole reduction associated with $l$ is a world line in such a congruence, i.e.,
$l$ is a principal null direction of the spacetime,
then the set of equations (\ref{eqsspinP_fin2}) simplifies to 
\begin{eqnarray}
\label{eqsspinP_as2}
\fl \frac{\rmd s}{\rmd \lambda} &=&0\ ,\quad 
\frac{\rmd S_2}{\rmd \lambda}  = \pi is -P^3\ ,\nonumber \\
\fl \frac{\rmd P^1}{\rmd \lambda} &=&  -\pi\bar P^3 - \bar \pi P^3  
                +i  s (\psi_2-\bar \psi_2)\ ,\quad 
\frac{\rmd P^3}{\rmd \lambda}  =0\ .
\end{eqnarray}

The formal integration of these final equations is straightforward, leading to
$s(\lambda)=s(\lambda_0)$ and $P^3(\lambda)=P^3(\lambda_0)$ being constant along $l$ and
\begin{eqnarray}
\label{solD}
S_2(\lambda)&=& is (\lambda_0)\int^\lambda_{\lambda_0} \pi(\lambda ') \rmd \lambda '
  -P^3(\lambda_0)(\lambda -\lambda_0) + S_2(\lambda_0)\ , \nonumber \\
P^1(\lambda) &=& -\bar P^3(\lambda_0) \int^\lambda_{\lambda_0} \pi(\lambda ') \rmd \lambda '
 - P^3(\lambda_0)\int^\lambda_{\lambda_0} \bar \pi(\lambda ') \rmd \lambda ' \nonumber \\
&&+i  s(\lambda_0) \int^\lambda_{\lambda_0} \left[\psi_2(\lambda ')-\bar \psi_2(\lambda ')\right] 
     \rmd \lambda '+P^1(\lambda_0)\ .
\end{eqnarray}

In the special case $P^3(\lambda_0)=0$ (i.e., $P$ null and aligned with $l$), the solution (\ref{solD}) reduces to
\begin{eqnarray}
\label{solD0}
S_2(\lambda)&=& is(\lambda_0)\int^\lambda_{\lambda_0} \pi(\lambda ') \rmd \lambda '  + S_2(\lambda_0)\ , \nonumber \\
P^1(\lambda) &=& i  s(\lambda_0) \int^\lambda_{\lambda_0}\left[\psi_2(\lambda ') -\bar \psi_2(\lambda ')\right] 
         \rmd \lambda '+P^1(\lambda_0)\ .
\end{eqnarray}

Eqs.~(\ref{solD}) and (\ref{solD0}) suggest the presence of  a \lq\lq
helicity-connection coupling" term  in $S_2$, as well as a \lq\lq
helicity-curvature coupling" term in $P^1$. We will come back to these couplings when discussing explicit examples.

\section{Explicit examples in the type D case}

In each case we adopt a well-known NP frame with the above properties, assuming that the world line of the multipole reduction is an affinely parametrized integral curve of the principal null direction $l$. We investigate the principal null direction corresponding to outgoing radial null geodesics in the Schwarzschild case and those which are asymptotically radial outgoing null geodesics in the Kerr case, and those in  related spacetimes which are analogous to these. The ingoing null geodesics of this type can be treated similarly.

\subsection{The Schwarzschild case}

Consider the Schwarzschild solution \cite{kraetal}  written in standard spherical coordinates
\beq
\fl\quad
\rmd s^2= \left( 1-\frac{2M}{r}\right)\rmd t^2 - \left( 1-\frac{2M}{r}\right)^{-1}\rmd r^2 
           - r^2 (\rmd \theta^2 +\sin^2\theta \rmd \phi^2)\ .
\eeq
Introduce the NP frame adapted to the outgoing radial null geodesics
\begin{eqnarray}
\fl\quad
l&=& \left(1-\frac{2M}{r}\right)^{-1}\partial_t+\partial_r\ ,  \quad 
n= \frac12 \left[\partial_t- \left(1-\frac{2M}{r}\right) \partial_r \right]\ ,\nonumber \\
\fl\quad
m&=& \frac{1}{\sqrt{2}r}\left(\partial_\theta +\frac{i}{\sin \theta}\partial_\phi \right)\ ,
\end{eqnarray}
which has the following nonvanishing  spin coefficients
\begin{eqnarray}
\fl\quad
&&\rho=-\frac1r\ , \quad 
\mu=-\frac{1}{2r}\left(1-\frac{2M}{r}\right)\ , \quad 
\gamma=\frac{M}{2r^2}\ ,\quad 
\beta=-\alpha = \frac{\sqrt{2}}{4r}\cot \theta\ ,
\end{eqnarray}
and the single nonvanishing Weyl scalar $\psi_2=-M/r^3$.
The integral curves of $l$ can be expressed in the form 
\beq\fl\quad
t=r-r_0+2M\ln \frac{r-2M}{r_0-2M}+t_0\ , \quad 
r=\lambda-\lambda_0+r_0\ , \quad 
\theta=\theta_0\ , \quad 
\phi=\phi_0\ ,
\eeq
where  $t_0, r_0, \theta_0, \phi_0$ are arbitrary integration constants taken at $\lambda=\lambda_0$. Then
Eqs.~(\ref{eqsspinP}) reduce to
\begin{eqnarray}
\frac{\rmd s}{\rmd \lambda} =0\ , \quad 
\frac{\rmd S_2}{\rmd \lambda} = -P^3\ , \quad
\frac{\rmd P^1}{\rmd \lambda}=0\ , \quad 
\frac{\rmd P^3}{\rmd \lambda} =0
\end{eqnarray}
with solution $s(\lambda)=s(\lambda_0)$, $S_2(\lambda)=S_2(\lambda_0)-P^3(\lambda_0)\,(\lambda-\lambda_0)$, $P^1(\lambda)=P^1(\lambda_0)$ and $P^3(\lambda)=P^3(\lambda_0)$.
The case $P^3(\lambda)=0$ is clearly trivial.

\subsection{The Taub-NUT case}

The Taub-NUT solution \cite{NUT,mistau,DemNew,Miller} is often written in the form
\begin{eqnarray}\fl\quad
\rmd s^2&=&\frac{\Delta_{\rm TN}}{r^2+\ell^2}(\rmd t+2\ell \cos \theta \rmd \phi)^2
           -(r^2+\ell^2)\left[\frac{\rmd r^2}{\Delta_{\rm TN}}
           +\rmd \theta^2+\sin^2\theta \rmd \phi^2\right]\ ,
\end{eqnarray}
where $\Delta_{\rm TN}=r^2-2Mr-\ell^2$, and it represents the spacetime of a black hole with an additional gravitomagnetic monopole charge associated with the parameter $\ell$.

Introduce the NP frame generalizing the previous Schwarzschild situation to which it reduces when $\ell=0$
\begin{eqnarray}
l&=& \frac{r^2+\ell^2}{\Delta_{\rm TN}}\partial_t+\partial_r\ ,  \quad 
n= \frac12 \left[\partial_t- \frac{\Delta_{\rm TN}}{r^2+\ell^2} \partial_r \right]\ ,\nonumber \\
m&=& \frac{1}{\sqrt{2}(r+i\ell)} \left[-2i\ell\cot \theta \partial_t + \partial_\theta 
              + \frac{i}{\sin \theta}\partial_\phi \right]\ ,
\end{eqnarray}
with nonvanishing spin coefficients
\begin{eqnarray}
\fl\quad
&&\rho=-\frac1{r-i\ell}\ , \quad 
\mu=-\frac12\frac{\Delta_{\rm TN}}{(r+i\ell)(r-i\ell)^2}\ , \quad 
\gamma=-\frac{i}2\frac{\ell+iM}{(r-i\ell)^2}\ ,\nonumber\\
&&\beta=-\bar\alpha = \frac{\sqrt{2}}{4}\frac{\cot \theta }{(r+i\ell)}\ ,
\end{eqnarray}
and the  single nonvanishing Weyl scalar $\psi_2=(\ell+iM)/(\ell+ir)^3$.
The integral curves of $l$ have the form
\begin{eqnarray}
\fl\quad t&=&r-r_0+M\ln \frac{r^2-2Mr-\ell^2}{r_0^2-2Mr_0-\ell^2} \nonumber \\
\fl\quad &&-2\sqrt{M^2+\ell^2}\left[ {\rm arctanh}\left(\frac{r-M}{\sqrt{M^2+\ell^2}}\right)
-{\rm arctanh}\left(\frac{r_0-M}{\sqrt{M^2+\ell^2}}\right) \right] + t_0\ , \nonumber \\
\fl\quad r&=&\lambda-\lambda_0+r_0\ ,\quad
\fl\quad \theta =\theta_0\ ,\quad
\fl\quad \phi =\phi_0\ ,
\end{eqnarray}
where  $t_0$, $r_0$, $\theta_0$ and $\phi_0$ are arbitrary integration constants taken at $\lambda=\lambda_0$.

The formal integration of (\ref{solD}) is then straightforward, leading to
$s(\lambda)=s(\lambda_0)$ and $P^3(\lambda)=P^3(\lambda_0)$ being constant along $l$ and
\begin{eqnarray}
\fl\quad S_2(\lambda)&=& -P^3(\lambda_0)(\lambda -\lambda_0) + S_2(\lambda_0)\ , \nonumber \\
\fl\quad P^1(\lambda) &=& \ell\, s(\lambda_0)\left\{ 
\frac{(\lambda-\lambda_0+r_0-M)^2-(M^2+\ell^2)}{[(\lambda-\lambda_0+r_0)^2+\ell^2]^2}
       -\frac{(r_0-M)^2-(M^2+\ell^2)}{(r_0^2+\ell^2)^2} \right\} 
\nonumber\\
\fl\quad & &  + P^1(\lambda_0)\ .
\end{eqnarray}

\begin{figure}
\typeout{*** EPS figure 1}
\begin{center}
\includegraphics[scale=0.45]{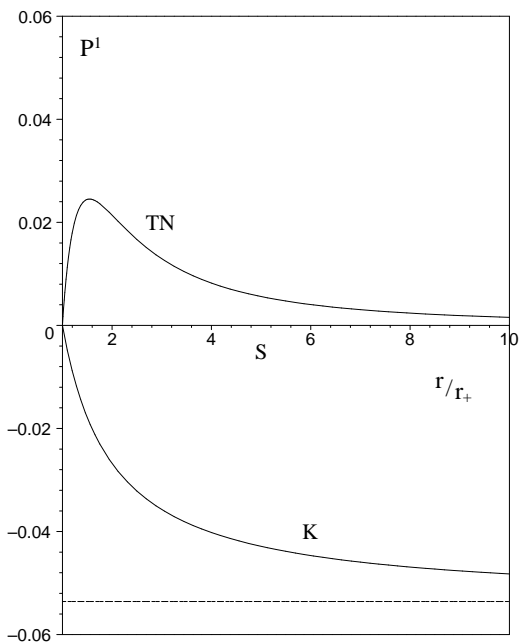}
\end{center}
\caption{
In the Schwarzschild, Taub-NUT and Kerr cases the varying component $P^1$ of the total 4-momentum is plotted versus the radial parameter $r/r_+$, where $r_+$ is the (outer) horizon in each of these spacetimes. The initial conditions are chosen so that $P^1(r_+)=0$ with $\lambda_0=r_0=r_+$ for all cases, $s(r_+)=1$ and $\ell/M=1$ for Taub-NUT, and $\theta_0=\pi/2$, Im$[P^3(r_+)]=1/(10\sqrt{2}a)$ and $a/M=.5$ for Kerr.
$P^1$ represents the component of the generalized momentum of the particle along the world line of the multipole reduction which is assumed to coincide with a principal null direction of the spacetime. In the Schwarzschild case this quantity is constant all the world line and identically zero for our choice of initial conditions. In the Taub-NUT case it varies but asymptotically returns to its initial value. In the Kerr case,   it varies and  does not return to its initial value, although it does approach the asymptotic value $-1/(10r_+)$, with our choice of initial conditions.
}
\label{fig:1}
\end{figure}

When $\lambda_0=r_0=r_+\equiv M+\sqrt{M^2+\ell^2}$ and $P^1(\lambda_0)=0$, i.e., the particle starts moving from the horizon $r_+$,
we have $\lambda=r$ and the solution is given by
$s(\lambda)=s(\lambda_0)$ and $P^3(\lambda)=P^3(\lambda_0)$ constant along $l$ and
\begin{eqnarray}
\fl\quad S_2(r)=S_2(r_+) -P^3(r_+)(r -r_+) \ , \qquad 
P^1(r) = \ell\, s(r_+)\,\frac{\Delta_{\rm TN}}{(r^2+\ell^2)^2} \ .
\end{eqnarray}
If in addition $P^3(r_+)=0$, then $S_2(r)=S_2(r_+)$ and $P$ is null and aligned with $l$
\beq
P=\ell\, s(r_+)\,\frac{\Delta_{\rm TN}}{(r^2+\ell^2)^2}\, l\ .
\eeq
The behavior of the varying component $P^1$ of the total 4-momentum
as a function of the radial parameter is shown in Fig.~\ref{fig:1}. 
Moving from $r_+$ out to infinity the proportionality factor with $l$ passes from zero to a maximum positive value at the real solution of $r^3-3Mr^2-3r\ell^2 +M\ell^2=0$ and then 
returns asymptotically to zero. 


\subsection{The C metric case}

The C metric solution \cite{kin69,kinwal,Farh}, which represents the spacetime associated with a uniformly accelerating black hole, is often written in the form
\beq
\fl\quad
\rmd s^2=H \rmd u^2 +2\rmd u \rmd r-2Ar^2\sin^2\theta \rmd r \rmd \theta 
     -\frac{r^2\sin^2\theta }{G}\rmd \theta^2-r^2G \rmd \phi^2\ ,
\eeq
where 
\begin{eqnarray}
H&=&1-\frac{2M}{r}-A^2r^2G-2A\cos \theta [r-3M(1-Ar\cos \theta)]\ , \nonumber \\
G&=&\sin^2\theta-2AM\cos^3\theta\ .
\end{eqnarray}
We limit our considerations to the  range of parameters $M<L_A\equiv 1/(3A\sqrt{3})$ where the C metric is static \cite{bcm}.
Introduce the NP frame
\begin{eqnarray}
l&=& \frac{2}{H}\partial_u-\partial_r,  \quad n= \frac{H}2 \partial_r \ ,\nonumber \\
m&=& -\frac{Ar\sqrt{G}}{\sqrt{2}}\partial_r -\frac{\sqrt{G}}{\sqrt{2}r\sin \theta}\partial_\theta 
       +\frac{i}{\sqrt{2}r\sqrt{G}}\partial_\phi\ ,
\end{eqnarray}
with nonvanishing spin coefficients
\begin{eqnarray}
\fl\quad
&&\rho=-\frac1{r}\ , \quad 
\mu=-\frac{H}{2r}\ , \quad 
\pi=-\frac{A\sqrt{G}}{\sqrt{2}}=-\tau\ , 
\quad \beta=- \frac{\sqrt{2}}{4}\frac{\cos \theta(1+3AM\cos \theta)}{r\sqrt{G}}\ , \nonumber\\
\fl\quad
&& \alpha=\frac{\sqrt{2}}{4r\sqrt{G}}[2ArG+\cos \theta(1+3AM\cos \theta)]\ , \nonumber\\
\fl\quad
&& \gamma=\frac{A^2r^3G+MA^2r^2\cos^2\theta(A\cos \theta r-3) + M - A\cos \theta r^2}{2r^2}
\end{eqnarray}
and Weyl scalar $\psi_2=-M/r^3$.

In order to write down the equations for the integral curves of $l$,
it is convenient to introduce the new variable
\beq
W= \frac{Ar}{1-Ar\cos \theta}\ ,
\eeq
in place of $r$ ($\theta$ is constant along $l$),
which has proven to be helpful in studying the horizons of the C metric, i.e., the solutions of the equation $H=0$, equivalent to
\beq
\label{eqW}
W^3-W+2MA=0\ .
\eeq
In the case $M<L_A$ its solutions are
\beq
W_1=2{\hat U}\ , \qquad 
W_2=-{\hat U}+\sqrt{3}{\hat V}\ , \qquad 
W_3=-{\hat U}-\sqrt{3}{\hat V}\ ,
\eeq
where
\beq
{\hat U}+i{\hat V}=\frac{1}{\sqrt{3}} \left(-\frac{M}{L_A} +i \sqrt{1-\frac{M^2}{L_A^2}} \right)^{1/3}\ ,
\eeq
i.e.,
\beq\fl\quad
{\hat U}=-\frac{1}{\sqrt{3}}\cos\left(\frac13{\rm arccos}\frac{M}{L_A}\right)\ , \quad 
{\hat V}=\frac{1}{\sqrt{3}}\sin\left(\frac13{\rm arccos}\frac{M}{L_A}\right)\ .
\eeq
In terms of these quantities one finds
\begin{eqnarray}
\fl u&=& u_0 + \frac{W_1}{(W_1-W_2)(W_1-W_3)}\ln \left|\frac{W-W_1}{W_0-W_1}\right| \nonumber \\
\fl && -\frac{W_2}{(W_1-W_2)(W_2-W_3)}\ln \left| \frac{W-W_2}{W_0-W_2}\right|
           +\frac{W_3}{(W_2-W_3)(W_1-W_3)}\ln\left| \frac{W-W_3}{W_0-W_3}\right|\ ,\nonumber \\
\fl r&=& -\lambda +\lambda_0+r_0, \quad \theta=\theta_0, \quad \phi=\phi_0\ ,
\end{eqnarray}
where
\beq
W_0=\frac{Ar_0}{1-Ar_0\cos \theta}\ .
\eeq

Finally, the formal integration of the equations of motion is straightforward, giving
$s(\lambda)=s(\lambda_0)$ and $P^3(\lambda)=P^3(\lambda_0)$ constant along $l$ and
\begin{eqnarray}
S_2(\lambda)&=& i \pi_0 s(\lambda_0)(\lambda-\lambda_0)  
      -P^3(\lambda_0)(\lambda-\lambda_0) + S_2(\lambda_0)\ , \nonumber \\
P^1(\lambda) &=& -2\pi_0(\lambda-\lambda_0){\rm Re}[ P^3(\lambda_0)]+P^1(\lambda_0)\ ,
\end{eqnarray}
where in this case $\pi$, depending only on $\theta$, is constant along $l$.

In the equatorial plane ($\theta=\pi/2$) we have $\pi_0=-A/\sqrt{2}$. Thus
\begin{eqnarray}
S_2(\lambda)&=& -\frac{iA}{2 \sqrt{2}} s(\lambda_0)(\lambda-\lambda_0)  
-P^3(\lambda_0)(\lambda-\lambda_0) + S_2(\lambda_0)\ , \nonumber \\
P^1(\lambda) &=& \sqrt{2}A(\lambda-\lambda_0){\rm Re}[ P^3(\lambda_0)]+P^1(\lambda_0)\ .
\end{eqnarray}

\subsection{The Kerr case}

Consider the Kerr metric \cite{kraetal} expressed in standard Boyer-Lindquist coordinates
\begin{eqnarray}
\fl \quad \rmd s^2&=&\frac{\Delta}{\Sigma}(\rmd t-a\sin^2\theta \rmd \phi)^2
-\frac{\sin^2\theta}{\Sigma}[(r^2+a^2)\rmd \phi -a \rmd t]^2-\frac{\Sigma}{\Delta}\rmd r^2-\Sigma \rmd \phi^2\ ,
\end{eqnarray}
where the standard notation  $\Delta=r^2+a^2-2Mr$, $\Sigma=r^2+a^2\cos^2\theta$ has been adopted.
Introduce the Kinnersley frame
\begin{eqnarray}
\fl\quad
l&=& \frac{1}{\Delta}\left[ (r^2+a^2)\partial_t +\Delta \partial_r + a \partial_\phi\right]\ ,\quad
n = \frac{1}{\Delta}\left[ (r^2+a^2)\partial_t -\Delta \partial_r + a \partial_\phi\right]\ , \nonumber \\
\fl\quad
m&=& \frac{1}{\sqrt{2}\sin \theta (r+ia\cos \theta)} \left(ia\sin^2\theta \partial_t 
 + \sin \theta \partial_\theta+i\partial_\phi \right)\ .
\end{eqnarray}

The nonvanishing  spin coefficients are
\begin{eqnarray}
\fl\quad
&&\rho=-\frac{1}{r-ia\cos \theta}\ , \quad 
\mu=-\frac12 \frac{\Delta}{(r-ia\cos \theta)\Sigma}\ , \quad
\tau=-\frac{ia\sin \theta}{\sqrt{2}\Sigma}\ , 
\nonumber \\
\fl \quad && 
\gamma=-\frac{i[Mi(r+ia\cos \theta)+a(r\cos \theta-ia)]}{2\Sigma (r-ia\cos \theta)}\ ,\nonumber\\
\fl \quad &&
\alpha =-\frac{1}{2\sqrt{2}}\frac{\cos \theta(r+ia\cos \theta)-2ia}{\sin \theta(r-ia\cos \theta)^2}\ , \nonumber \\
\fl \quad &&  
\pi=\frac{ia\sin \theta}{\sqrt{2}(r-ia\cos \theta)^2}\ ,\quad 
\beta=\frac{\cos \theta}{2\sqrt{2}\sin \theta(r+ia\cos \theta)}\ ,
\end{eqnarray}
and the Weyl scalar is $\psi_2=-M/(r-ia\cos \theta)^3$.
The integral curves of  $l$ are
\begin{eqnarray}
\fl &&t=t_0+(r-r_0)+M\ln \left(\frac{r^2-2Mr+a^2}{r_0^2-2Mr_0+a^2}\right)\nonumber \\
\fl && \qquad -\frac{2M}{\sqrt{M^2-a^2}}\left[
{\rm arctanh} \left( \frac{r-M}{\sqrt{M^2-a^2}}\right)
-{\rm arctanh} \left( \frac{r_0-M}{\sqrt{M^2-a^2}}\right)\right]\ , \nonumber \\
\fl && r=r_0+\lambda-\lambda_0, \quad
 \theta=\theta_0\ , \nonumber \\
\fl && \phi=\phi_0-\frac{a}{\sqrt{M^2-a^2}}\left[{\rm arctanh}  
\left( \frac{r-M}{\sqrt{M^2-a^2}}\right) -{\rm arctanh}  
\left( \frac{r_0-M}{\sqrt{M^2-a^2}}\right)\right]
\end{eqnarray}
and represent outgoing null geodesics which are asymptotically radial far from the horizon.
Eqs.~(\ref{eqsspinP})  then have the following solution
\begin{eqnarray}
\label{solKerr}
\fl\quad   s(\lambda)&=&s(\lambda_0)\ , \nonumber \\
\fl \quad  S_2(\lambda)&=&S_2(\lambda_0)-P^3(\lambda_0)(\lambda-\lambda_0) \nonumber \\
\fl\quad && +\frac{1}{\sqrt{2}}a\sin \theta s(\lambda_0)
\left[ \frac{1}{\lambda-\lambda_0+r_0-ia\cos\theta}-\frac{1}{r_0-ia\cos \theta}\right]\ ,\nonumber \\
\fl\quad   P^1(\lambda)&=&P^1(\lambda_0)+\frac{1}{\sqrt{2}}ia\sin \theta\left\{
\bar P^3(\lambda_0) \left[ \frac{1}{\lambda-\lambda_0+r_0-ia\cos \theta}-\frac{1}{r_0-ia\cos \theta}\right] \right.\nonumber \\
\fl\quad &&\left. - P^3(\lambda_0) \left[ \frac{1}{\lambda-\lambda_0+r_0+ia\cos \theta}-\frac{1}{r_0+ia\cos \theta}\right]
\right\}
\nonumber \\
\fl\quad && -2aM\cos \theta s(\lambda_0)\left[  \frac{\lambda-\lambda_0+r_0}{(\lambda-\lambda_0+r_0)^2+a^2\cos^2\theta}-\frac{r_0}{r_0^2+a^2\cos^2\theta}\right]\ ,
\end{eqnarray}
where $\sin \theta$ and $\cos \theta$ refer to the fixed value $\theta=\theta_0$. 

For example, on the equatorial plane $\theta_0=\pi/2$, Eq.~(\ref{solKerr}) reduces to
\begin{eqnarray}
\label{solKerreq}
\fl\quad   s(\lambda)&=&s(\lambda_0)\ , \nonumber \\
\fl \quad  S_2(\lambda)&=&S_2(\lambda_0)-P^3(\lambda_0)(\lambda-\lambda_0)
+\frac{1}{\sqrt{2}}as(\lambda_0)\left( \frac{1}{\lambda-\lambda_0+r_0}-\frac{1}{r_0}\right)\ ,\nonumber\\
\fl\quad   P^1(\lambda)&=&P^1(\lambda_0)+\sqrt{2}a  {\rm Im}[P^3(\lambda_0)] \left( \frac{1}{\lambda-\lambda_0+r_0} 
  -\frac{1}{r_0}\right)\ .
\end{eqnarray}
In the special case $r_0=\lambda_0=r_+\equiv M+\sqrt{M^2-a^2}$ and $P^1(\lambda_0)=0$, i.e.,the particle starts moving from the horizon $r_+$, we have $\lambda=r$ and
\begin{eqnarray}
\label{solKerreq_spec}
\fl\quad   s(r)&=&s(r_+)\ , \quad
  S_2(r) = S_2(r_+)-P^3(r_+)(r-r_+)+\frac{1}{\sqrt{2}}as(r_+)\left( \frac{1}{r}-\frac{1}{r_+}\right)\ ,\nonumber\\
\fl\quad   P^1(r)&=&\sqrt{2}a  {\rm Im}[P^3(r_+)]\left( \frac{1}{r}-\frac{1}{r_+}\right)\ .
\end{eqnarray}
If in addition $P^3(r_+)=0$, then we have
\begin{eqnarray}
\label{solKerr_eq_P30}
\fl\quad  && s(r)=s(r_+)\ ,\quad
 S_2(r)=S_2(r_+)+\frac{1}{\sqrt{2}}as(r_+)\left( \frac{1}{r}-\frac{1}{r_+}\right)\ .
\end{eqnarray}
We see that $S_2$ (carrying spin weight $-1$) is coupled (via the Kerr rotational parameter $a$) to $S_1$ (i.e.,$s$, with spin weight $0$): in a sense, a helicity-rotation coupling factor $as$ is necessary to ensure spin-weight conservation on both sides of the equation for $S_2$. 

The behaviour of the varying component $P^1$ of the total 4-momentum 
as a function of the radial parameter is shown in Fig.~\ref{fig:1}. $P^1$ is always negative in this case, 
and goes to the asymptotic value $-\sqrt{2}a  {\rm Im}[P^3(r_+)]/r_+$ for $r\to\infty$. 
Note that in the Schwarzschild limit ($a=0$) $P^1$ vanishes identically.

\subsection{The Kasner case}

The Kasner metric \cite{kraetal} representing an anisotropic universe model is
\beq
\rmd s^2= \rmd t^2 + t^{2p_1}\rmd x^2+ t^{2p_2}\rmd y^2+ t^{2p_3}\rmd z^2\ ,
\eeq
with $p_1+p_2+p_3=p_1^2+p_2^2+p_3^2=1$. Choosing Kasner index values $p_1=-1/3$, $p_2 = p_3 = 2/3$ (and permutations) one has the
Petrov type $D$ case,  and introducing the NP tetrad
\begin{eqnarray}
l&=&\frac{t^{-p_1}}{\sqrt{2}}[\partial_t+t^{-p_1}\partial_x]\,, \
n=\frac{t^{p_1}}{\sqrt{2}}[\partial_t-t^{-p_1}\partial_x]\,,\
m=\frac1{\sqrt{2}}[t^{-p_2}\partial_y+it^{-p_3}\partial_z]\,,
\end{eqnarray}
one finds the nonzero spin coefficients
\beq
\gamma=\frac{\sqrt{2}}{6}t^{-2p_2}=\frac{\mu}2\ , \quad 
\rho=-\frac{\sqrt{2}}3 t^{-p_2}\ ,
\eeq
while the only surviving Weyl scalar is $\psi_2=-2/(9t^2)$.
The integral curves of  $l$ are
\begin{eqnarray}
t&=&\left[\frac{\sqrt{2}}{3}(\lambda-\lambda_0)+t_0^{2/3}\right]^{3/2}\ , \quad
x = x_0+\frac16(\lambda-\lambda_0)^2+\frac{t_0^{2/3}}{\sqrt{2}}(\lambda-\lambda_0)\ ,\nonumber\\
y&=&y_0\ ,\quad
z = z_0\ ,
\end{eqnarray}
and
$s(\lambda)=s(\lambda_0)$, $P^1(\lambda)=P^1(\lambda_0)$, $P^3=P^3(\lambda_0)$, 
$S_2=-P^3(\lambda_0)(\lambda-\lambda_0)+S_2(\lambda_0)$.

\section{Explicit example in the type N case}

Consider the metric associated with a single gravitational wave \cite{kraetal} in coordinates $u,v,x,y$
\beq
\rmd s^2=-{\mathcal H}\rmd u^2+\rmd u \rmd v -\rmd x^2-\rmd y^2\ ,
\eeq
where
\beq
{\mathcal H}=f_+(u)(x^2-y^2)+2f_\times (u) xy\ ,
\eeq
and $f_{+, \times }$ represent the two possible polarization states of the wave.
The natural NP tetrad is
\beq
l=2\partial_v\ , \quad 
n=\partial_u + {\mathcal H}\partial_v\ , \quad 
m=\frac{1}{\sqrt{2}}(\partial_x+i\partial_y)
\eeq
with the single nonvanishing spin coefficient
\beq
\nu=-\frac{1}{\sqrt{2}}\left[ f_+(u)(x+iy)+if_\times (u) (x-iy)\right]
\eeq
and Weyl scalar $\psi_4=-[f_+(u)+if_\times (u)]$.
The integral curves of $l$ are
\beq\fl\quad
u=u_0, \quad v=2(\lambda-\lambda_0)+v_0,\quad x=x_0, \quad y=y_0,
\eeq
with  $u_0, v_0, x_0, y_0$ arbitrary integration constants taken at $\lambda=\lambda_0$.
The Mathisson-Papapetrou equations for the world line of the multipole reduction chosen within the congruence $l$ are
\begin{eqnarray}
\label{eqsspinP_N}
\frac{\rmd s}{\rmd \lambda}=0\ , \quad
\frac{\rmd S_2}{\rmd \lambda}=-P^3\ ,\quad
\frac{\rmd P^1}{\rmd \lambda} =0\ ,\quad
\frac{\rmd P^3}{\rmd \lambda} =0\ ,
\end{eqnarray}
with solution: $s(\lambda)=s(\lambda_0)$, $S_2(\lambda)=S_2(\lambda_0)-P^3(\lambda_0)\,(\lambda-\lambda_0)$, 
$P^1=P^1(\lambda_0)$ and $P^3=P^3(\lambda_0)$.
The case $P^3=0$ is trivial.

\section{Null circular orbits in some familiar type D spacetimes}

Instead of choosing $l$ to be a (geodesic) principal null direction of the background spacetime, one can allow it to be the tangent vector to any other null geodesic.
Null circular geodesics in some of the above type D spacetimes are of special interest; in the black hole case they represent the last possible circular orbits as one approaches the horizon.
To describe this case,
Eqs.~(\ref{eqsspinP_fin}) reduce to
\begin{eqnarray}
\frac{\rmd s}{\rmd \lambda} &=&0\ ,\nonumber \\
\frac{\rmd S_2}{\rmd \lambda}  &=& \pi_0 is -P^3\ ,\nonumber \\
\frac{\rmd P^1}{\rmd \lambda} &=& -\pi_0\bar P^3 - \bar \pi_0 P^3  +i  s\, (\psi_2-\bar \psi_2)\ ,\nonumber \\
\frac{\rmd P^3}{\rmd \lambda}  &=& \bar \psi_0 \bar S_2\ ,
\end{eqnarray}
where we have set $\epsilon=0$, assuming $l$ to be affinely parametrized, and in addition $\psi_1=0$, because this is the case in all the explicit examples considered below and we have used the property that all of the nonvanishing spin coefficients ($\pi_0$ denotes the constant value of $\pi$ when evaluated along the orbit) and the Weyl scalars are constant along the orbit. Then the system can be easily integrated and the solution is 
\begin{eqnarray}
\fl s(\lambda) &=& s(\lambda_0)\ ,\nonumber \\
\fl P^1(\lambda) &=& 2{\rm Re}\{\pi_0[\bar S_2(\lambda)-\bar S_2(\lambda_0)]\}
    -2s(\lambda_0)\,{\rm Im} [\psi_2(\lambda_0)](\lambda-\lambda_0) +P^1(\lambda_0)\ ,\nonumber \\
\fl P^3(\lambda)  &=& \pi_0 is(\lambda_0)-\frac{\rmd S_2}{\rmd \lambda}\ ,
\end{eqnarray}
where $S_2(\lambda)$ satisfies the following equation
\beq
\frac{\rmd^2 S_2}{\rmd \lambda^2}=-\bar\psi_0 \bar S_2\ .
\eeq
For example, in the cases in which $\psi_0$ (constant along the orbit) is real and positive, the solution for $S_2$ is given by 
\beq\fl\quad
S_2(\lambda) = \left[c_1\cos{\sqrt{\psi_0}\lambda}+c_2\sin{\sqrt{\psi_0}\lambda}\right]
              +i\left[c_3e^{\sqrt{\psi_0}\lambda}+c_4e^{-\sqrt{\psi_0}\lambda}\right]\ .
\eeq
When instead $\psi_0$ is not purely real, we have to solve a constant coefficient second order system of coupled  ordinary differential equations for the real and imaginary parts of $S_2$, which is also trivial to do.
We give the essential details to obtain the solution for the Kerr, Schwarzschild, Taub-NUT and C metric spacetimes in the following subsections.

\subsection{Kerr and Schwarzschild spacetimes}

In the case of the Kerr solution, it is well known that in the equatorial plane ($\theta=\pi/2$) there exist two circular null geodesics at radii 
\beq
r_{\rm (geo), \pm }=2M\left\{ 1+\cos\left[ \frac23 {\rm arccos} \left( \pm \frac{a}{M}\right) \right] \right\}
\eeq
which are real solutions of the cubic equation
\beq
r^3-6Mr^2+9M^2r-4Ma^2=0\ .
\eeq
To discuss this case, we consider the  following NP frame field adapted to generic co-rotating and counter-rotating circular null orbits
\begin{eqnarray}
\fl\quad
l_\pm &=& \partial_t + \frac{2aMr\sin \theta \pm\sqrt{\Delta}\Sigma}{\sin \theta [(r^2+a^2)\Sigma +2a^2Mr\sin^2 \theta]} \partial_\phi \ ,\nonumber \\
\fl\quad
n_\pm &=& \frac{(r^2+a^2)\Sigma +2a^2Mr\sin^2 \theta}{2\Sigma \Delta}\left[ \partial_t + \frac{2aMr\sin \theta \mp\sqrt{\Delta}\Sigma}{\sin \theta [(r^2+a^2)\Sigma +2a^2Mr\sin^2 \theta]} \partial_\phi\right]\ ,\nonumber \\
\fl\quad
m &=& \frac{1}{\sqrt{2\Sigma}}\left( \sqrt{\Delta}\partial_r +i \partial_\theta\right)\ .
\end{eqnarray}
The congruence $l$ contains the null geodesic orbits in the equatorial plane. 
Along such orbits, $\theta=\pi/2$ and $r=r_{\rm (geo), \pm }$ and the only nonvanishing spin coefficients reduce to
\begin{eqnarray}
\fl\quad && \pi=-\tau=\frac{\sqrt{2}(r-M)}{4r\sqrt{\Delta}}\ , \quad
\beta= \alpha + \frac{\sqrt{\Delta}}{\sqrt{2}r^2}\ , \nonumber \\
\fl\quad && \alpha= -\frac{\sqrt{2}}{4r^2\sqrt{\Delta}} 
  \frac{(r^2+a^2)(r^3-3Mr^2+ra^2+Ma^2)\pm aM\sqrt{\Delta}(3r^2+a^2)} {r^3+a^2r+2Ma^2}\ , 
\end{eqnarray}
and the nonvanishing Weyl scalars are
\begin{eqnarray}
\psi_0 &=&-\frac{3M\Delta}{r^3(r^3+a^2r+2a^2M)} 
\left[r^2+a^2\mp a\sqrt{\Delta} \right]^2\ ,\quad
\psi_2 = \frac{M}{2r^3}\ , \nonumber \\
\psi_4&=& -\frac{3M}{4r^5\Delta} \left[r^2+a^2\pm a\sqrt{\Delta} \right]^2\ ,
\end{eqnarray}
where $r=r_{\rm (geo), \pm }$ is assumed. 
Thus the spin coefficients and Weyl scalars are all constant along the orbit.
From Eqs.~(\ref{eqsspinP_fin}) we have
\begin{eqnarray}
\label{eqsspinP_fin_new}
& \frac{\rmd s}{\rmd \lambda} =0\ , \quad  
  \frac{\rmd S_2}{\rmd \lambda}  = \pi is -P^3\ ,\nonumber \\
& \frac{\rmd P^1}{\rmd \lambda} = -\pi( \bar P^3 + P^3)\ ,\quad
 \frac{\rmd P^3}{\rmd \lambda}  = \psi_0 \bar S_2\ .
\end{eqnarray}
which can be easily integrated because all the coefficients are constant.

In the limit of the Schwarzschild spacetime all these results simplify. The NP tetrad adapted to the null circular orbits is
\begin{eqnarray}
l_\pm &=& \partial_t \mp\frac{\sqrt{r(r-2M)}}{r^2\sin \theta} \partial_\phi
\ ,\quad
n_\pm = \frac{r}{2(r-2M)}\left[ \partial_t \pm \frac{\sqrt{r(r-2M)}}{r^2\sin \theta} \partial_\phi\right]
\ ,\nonumber \\
m &=& \frac{1}{\sqrt{2}}\left( \sqrt{1-\frac{2M}{r}}\partial_r +\frac{i}{r} \partial_\theta\right)
\end{eqnarray}
Along the two equatorial null geodesics  orbits $\theta=\pi/2$ and $r=3M$
the only nonvanishing spin coefficients are
\beq
\pi=\beta=-\tau=-\frac{\sqrt{6}}{18M}\ ,
\eeq
while the nonzero Weyl scalars
\beq
\psi_0=-\frac{1}{27M^2}\ , \quad 
\psi_2=\frac{1}{54M^2}\ , \quad 
\psi_4=-\frac{1}{12M^2}\ ,
\eeq
are all constant.
From Eqs.~(\ref{eqsspinP_fin_new}) we  easily get the solutions.

\subsection{Taub-NUT spacetime}
\label{TNcirc}

Introduce an NP frame adapted to a generic circular null orbit
\begin{eqnarray}\fl\quad
l&=& \partial_t-\frac{\sqrt{\Delta_{\rm TN}}} 
  {(r^2+\ell^2)\sin\theta+2\ell\sqrt{\Delta_{\rm TN}}\cos\theta}\partial_r\ ,  \nonumber\\
\fl\quad
n&=& \frac{(r^2+\ell^2)^2\sin^2\theta-4\ell^2\Delta_{\rm TN}\cos^2\theta}
{2(r^2+\ell^2)\Delta_{\rm TN}\sin^2\theta}
\left[ \partial_t+\frac{\sqrt{\Delta_{\rm TN}}}
  {(r^2+\ell^2)\sin\theta-2\ell\sqrt{\Delta_{\rm TN}}\cos\theta}\partial_r\right]\ ,\nonumber \\
\fl\quad
m&=& \frac{\sqrt{2}}{2(r^2+\ell^2)^{1/2}}\left[\sqrt{\Delta_{\rm TN}}\partial_r+i\partial_\theta \right]\ .
\end{eqnarray}
Null geodesics exist off the equatorial plane satisfying the conditions
\begin{eqnarray}\fl\quad
\label{tnnullgeos}
r^3-3Mr^2-3\ell^2r+M\ell^2=0\ ,\qquad 
\cos^2\theta=\frac{4\ell^2\Delta_{\rm TN}}{(r^2+\ell^2)^2+4\ell^2\Delta_{\rm TN}}\ . 
\end{eqnarray}
The first equation gives
\beq
r_{\rm (geo)}=M+2\sqrt{M^2+\ell^2}\cos \left[ \frac13 {\rm arccos}\left( \frac{M}{\sqrt{M^2+\ell^2}}\right)\right]\ .
\eeq

The only nonvanishing  spin coefficients (evaluated along the orbits, i.e.,  using Eqs.~(\ref{tnnullgeos}) to simplify general formulas) therefore reduce to
\begin{eqnarray}\fl\quad
\pi&=&-\bar\tau=\frac{\sqrt{2}}{4(r^2+\ell^2)^{1/2}}\left[\frac{r-M}{\sqrt{\Delta_{\rm TN}}}-i\cot\theta \right]\ , \quad 
\nonumber\\
\fl\quad
\alpha&=&\bar\beta+\frac{\sqrt{2}}{2}\frac{r\sqrt{\Delta_{\rm TN}}}{(r^2+\ell^2)^{3/2}} \nonumber\\
\fl\quad
&=&i\frac{\sqrt{2}}{4}\frac{\ell\sqrt{\Delta_{\rm TN}}}
  {(r^2+\ell^2)^{3/2}\sin\theta}\frac{4\ell\sqrt{\Delta_{\rm TN}}
  (r^2+\ell^2)\cos\theta-[(r^2+\ell^2)^2+4\ell^2\Delta_{\rm TN}]\sin\theta}{(r^2+\ell^2)^2-4\ell^2\Delta_{\rm TN}}\ , \nonumber\\
\fl\quad
\nu&=&i\frac{\sqrt{2}}{8}\frac{(r^2+\ell^2)^2+4\ell^2\Delta_{\rm TN}}
  {\Delta_{\rm TN}(r^2+\ell^2)^{5/2}\sin^3\theta}[2\ell\sqrt{\Delta_{\rm TN}}\sin\theta+(r^2+\ell^2)\cos\theta]\ ,
\end{eqnarray}
and the nonvanishing Weyl scalars are
\begin{eqnarray}\fl\quad
\psi_0&=&3\frac{\Delta_{\rm TN}(\ell-iM)(\ell+ir)^3}{[(r^2+\ell^2)^4-16\ell^4\Delta_{\rm TN}^2]^2}
  \Big\{(r^2+\ell^2)^4+16\ell^4\Delta_{\rm TN}^2\nonumber\\
\fl\quad
&&+4\ell\sqrt{\Delta_{\rm TN}}(r^2+\ell^2)[(r^2+\ell^2)^2+4\ell^2\Delta_{\rm TN}]\sin\theta\cos\theta\Big\}\ , \nonumber\\
\fl\quad
\psi_2&=&\frac{i}{2}\frac{\ell-iM}{(r+i\ell)^3}\ , \quad
\psi_4 = -\frac{i}{4}\frac{[(r^2+\ell^2)^4-16\ell^4\Delta_{\rm TN}^2]^2}
  {\Delta_{\rm TN}^2(r^2+\ell^2)^3(\ell+ir)^6}\psi_0\ .
\end{eqnarray}
The integration of the equations of motion then follows easily.

\subsection{C metric}

Introduce the following  NP frame
\begin{eqnarray}
l&=&\frac{1}{\sqrt{2}}[\sqrt{H}\rmd u + \frac{\rmd r -Ar^2\sin \theta \rmd \theta}{\sqrt{H}}+r\sqrt{G}\rmd \phi]\ ,  \nonumber\\
n&=& \frac{1}{\sqrt{2}}[\sqrt{H}\rmd u + \frac{\rmd r -Ar^2\sin \theta \rmd \theta}{\sqrt{H}}-r\sqrt{G}\rmd \phi]\ ,\nonumber \\
m&=& \frac{1}{\sqrt{2}}\left[ \frac{\rmd r}{\sqrt{H}}+r\sin \theta \left(\frac{i}{\sqrt{G}}-\frac{Ar}{\sqrt{H}}\right)\right]\ .
\end{eqnarray}

Null geodesics exist in the equatorial plane ($\theta=\pi/2$) at $r=3M$.
Evaluated along this orbit, the only nonvanishing spin coefficients are:
\begin{eqnarray}
&& \beta= -\frac{i}{\sqrt{6}M}\left( \frac{M}{L_A}+i\sqrt{1-M^2/L_A^2}\right)
= - \bar \alpha 
= - \tau=\bar \pi\ ,
\end{eqnarray}
while the Weyl scalars reduce to
\beq
\psi_0=-\frac{1}{18M^2}\ , \quad 
\psi_2=\frac{1}{54 M^2}\ , \quad 
\psi_4=-\frac{1}{18M^2}\ .
\eeq
The equations of motion are easily integrated.

\section{Concluding remarks}

We have studied the motion of massless spinning test particles in algebraically special vacuum spacetimes according to the Mathisson-Papapetrou model supplemented by Pirani's conditions in the framework of the Newman-Penrose formalism. We have considered the case in which the particle follows a null geodesic world line belonging to a congruence of a principal null direction of the background spacetime, a condition which allows complete integration of the equations of motion and for this case
we have discussed explicit examples for both type D  spacetimes (Schwarzschild, Kerr, Taub-NUT, C metric, Kasner) and type N  spacetimes (gravitational wave), revealing helicity couplings with connection coefficients as well as the curvature of the background.
Finally, we have also considered for some familiar type D spacetimes the motion of a massless spinning test particle along a circular null geodesic, a case which can also be completely solved analytically.

\end{document}